\documentclass[twocolumn,trackchanges]{aastex701}
\usepackage[utf8]{inputenc}
\usepackage{amsmath}
\usepackage{amssymb}
\usepackage{mathtools}
\usepackage{wasysym}
\usepackage{graphicx}
\usepackage{listings}
\usepackage{soul}
\usepackage{textcomp}
\usepackage{natbib}
\usepackage{leftindex}
\usepackage{comment}
\usepackage{booktabs}   
\usepackage{multirow}
\newcommand{\unit}[1]{\,\mathrm{#1}}
\begin{document}
\title{On the Missing Red Giants near the Galactic Center}

\author[orcid=0009-0005-2562-7680,sname='North America']{Taeho Kim}
\affiliation{Princeton University, Department of Astrophysical Sciences}
\email[show]{tk3766@princeton.edu}  
\author[orcid=0000-0002-6710-7748,sname='North America']{Jeremy Goodman}
\affiliation{Princeton University, Department of Astrophysical Sciences}
\email[show]{jeremy@astro.princeton.edu}  

\begin{abstract}
\noindent{}There is a long-acknowledged deficiency of bright red giants relative to fainter old stars within a few arc seconds of Sgr A*.  We explore whether this could be due to tidal stripping by the central black hole. This requires putting the stars onto highly eccentric orbits, for which we evaluate diffusion by both scalar resonant and non-resonant relaxation of the orbital angular momentum. We conclude that tidal stripping does not discriminate sufficiently between main-sequence and red giant stars. While the tidal loss cone increases with stellar radius, the rate of diffusion into the loss cone increases only logarithmically, whereas the lifetime on the red giant branch decreases more rapidly than $R_*^{-1}$.  In agreement with previous studies, we find that stellar collisions are a more likely explanation for the deficiency of bright red giants relative to fainter ones.
\end{abstract}

\section{Introduction}\label{sec:intro}
\subsection{The Evidence}\label{subsec:evidence}
Working with the $2.3\,\mathrm{\mu m}$ CO bandhead emission seen in integrated light, \cite{selg90} first inferred a near absence of bright red giants from the innermost $0.6\,\mathrm{pc}$ of the Galactic Center (GC).\footnote{\cite{selg90} identified the GC with IRS~16, a complex of infrared sources 1-$2''$ from the radio source Sgr~A*. They took the distance to the GC as $8.5\unit{kpc}$, which is close enough to modern values (e.g., $R_0=8246\pm 9\unit{pc}$, \citealt{GRAVITY2020}) that the conversion from arcseconds to parsecs was nearly the same: $1\unit{pc}\approx25''$.} 
Adaptive optics has made it possible to count and characterize individual stars in this region \citep{Schoedel+2007,Do+2009,Buchholz+2009,Fritz+2016,cano24}.

The general upshot of these works is that the projected number density, $N(r)$, of late-type stars is roughly constant with respect to projected distance ($r$) from Sgr~A$^*$ at small radii, but falls as a power law at larger $r$. The value of the break radius varies among studies from $6''$ (e.g. \citealt{Buchholz+2009}) to $\sim 20''$ \citep{Fritz+2016}.
In particular, \cite{schodel20} counted stars down to an observed magnitude of 19 in the $K_s$ band, finding that the slope of the counts within the inner $4''$ depends on magnitude: stars fainter than $K_s=15$ continue inward as a power law, while the brighter counts flatten.
Most of the late-type stars are believed to be old ($\gtrsim 10\unit{Gyr}$), with masses close to that of the Sun \citep{schodel20}.
Famously however, there is a minority of early-type and apparently young stars in the inner few tenths of a parsec, whose surface density rises more steeply inward \citep{Ghez+2003,Paumard+2006,Lu+2008,Bartko+2009,Genzel+2010,Do+2013,Yelda+2014,cano24}.

The absolute $K_s$ magnitude of the Sun in the Vega system is 3.27 \citep{will18}. At the distance of the GC, and allowing for $2.62\ \unit{mag}$ of extinction at $K_s$ \citep{schodel20}, the corresponding apparent magnitude would be $K_s\approx20.3$.
Therefore, insofar as the late-type stars counted by \cite{schodel20} are old and therefore scarcely more massive than the Sun, they should all be post-main-sequence objects, even those as faint as $K_s=19$.
Allowing for redder-than-solar colors on the RGB, we estimate that the critical magnitude $K_s=15$ at which the counts flatten corresponds bolometrically to $\sim 65\,L_{\astrosun}$.
If one does not correct for the difference in color, one gets $135\pm20\,L_{\astrosun}$.
The latter range is used in our dynamical calculations, but it would not have changed our conclusions to have used the smaller luminosity.

\subsection{Previous Theoretical Work}
Several mechanisms have been explored to explain the projected deficit in red giants near the GC: collisions and close encounters with other stellar bodies and stellar mass black holes \citep{Genzel+1996, Bailey+Davies1999, atep99}, tidal encounters with the SMBH \citep{Davies+King2005}, collisions with a gaseous disk \citep{Amaro-Seoane+2020}, and an AGN jet \citep{Zajacek+2020,Kurfuerst+2024}. 
In the present paper, we simulate both collisional interactions and eccentricity diffusion through relaxation to observe their combined effects on the post-simulation distribution of red giants in the GC. 

Our approach resembles that of \cite[hereafter RSSI23]{rose23}, but with important differences.
RSSI23 simulated stellar clusters starting from a broad initial mass function.
They focused on the effects of collisions on the main sequence, each collision resulting in partial mass loss or a merger, followed by modified main-sequence evolution.
Stars were deleted upon reaching the end of the main-sequence.
RSSI23 allowed for two-body relaxation in orbital energy (but not eccentricity), finding it to be important only for the lower-mass stars because of their longer main-sequence lifetimes.
In agreement with previous work, RSSI23 concluded that collisions on the main sequence could substantially deplete the progenitors of red giants in the inner $0.1\unit{pc}$.

\begin{figure}
\figurenum{1}
\epsscale{1.18}
\plotone{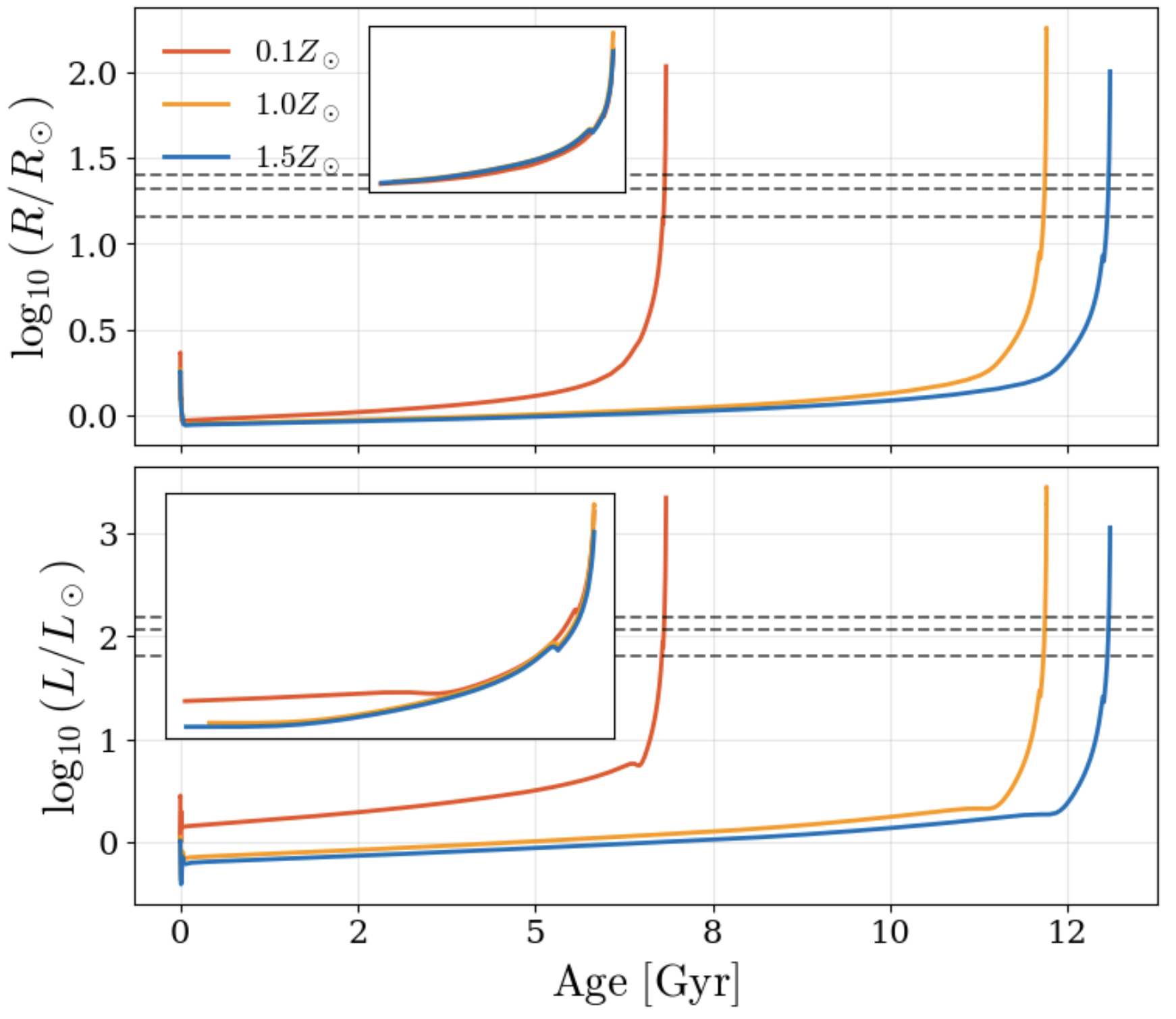}
\caption{Evolution in luminosity (\textit{upper panel}) and radius (\textit{lower panel}) of solar-mass stars with various metallicities as computed with MESA version 2024-03-1. 
Dashed lines denote the luminosity and radius corresponding to the $K_{s}\approx15$ (65$L_{\astrosun}$, 115$L_{\astrosun}$, 155$L_{\astrosun}$) for a solar-type red giant. 
The insets describe the last Gyr of luminosity/radius evolution for each metallicity for ease of comparison. 
}
\label{fig1}
\end{figure}

We don't dispute this conclusion but, as noted above, star counts indicate selective depletion of brighter red giants {\it relative to fainter red giants} (or at any rate, post-main-sequence objects) (\S\ref{subsec:evidence}).
Therefore, we study the collisional history of stars that are already on the RGB, taking into account the changes in their collisional cross sections as their radii expand.
Unlike RSSI23, we take all of our stars to have the same initial mass.
This is less of a restriction than it might seem because, first of all, apart from the effects of stellar mergers, only stars of initial masses close to that of the Sun have main-sequence lifetimes comparable to the age of the Galaxy, hence putting them on the RGB today; and secondly, because evolution on the RGB---in particular, radius and luminosity versus time---is insensitive to initial total masses $\lesssim 2\,M_{\astrosun}$, and also to metallicity, being governed instead by the growing mass of a degenerate helium core (Figs.~\ref{fig1} \& \ref{fig2}; \citealp{Salaris+2002}).

We also consider tidal stripping of red giants by the central supermassive black hole, in more detail than \cite{Davies+King2005}.\footnote{Our naive initial motivation for this was that, in a homogeneous stellar system and in the gravitational-focusing regime, the cross-section for tidal stripping scales with the masses of individual field stars as $m_{f}^{4/3}$, whence one might expect the SMBH to dominate.  But the stellar cluster is not homogeneous, the near-keplerian orbital structure matters, and---for the brighter red-giant test stars at least---the geometric cross section dominates in collisions with other stars.}
For orbital semimajor axes $\sim0.1\unit{pc}$, such stripping can occur only on near-radial orbits, so we model diffusion of orbital eccentricity due to both conventional (nonresonant) two-body relaxation and to resonant relaxation.
\cite{balex16} have previously concluded that resonant relaxation (which can significantly contribute to the loss-cone diffusion when compared to non-resonant relaxation in certain phase-space regions, see Section \ref{sec2}) occurs over a significantly longer timescale than the lifespan of the red giant. However, their studies focused on a collision-free test star with fixed characteristics, while we evolve its radius and include collisions.

\subsection{Paper Overview}
\S\ref{sec2} describes our treatment of diffusion of orbital eccentricity under the influence of both resonant and non-resonant relaxation.  For the latter, we compute diffusion coefficients with the publicly available code \texttt{scRRpy} \citep{bar19}.
The rate at which stars enter the tidal-stripping loss cone is first estimated by solving a steady-state backward Fokker-Planck equation based on the (resonant and non-resonant) diffusion coefficients.
This formalism, however, does not allow for the evolution of stellar radius---and hence of the loss cone---on the RGB, so we introduce a more flexible Monte-Carlo method, which allows us to follow the expanding radius and include collisions.
In \S\ref{sec3}, we add stellar collisions to the latter method.  Orbit-averaged collision rates are derived in Appendix~A for an isotropic, power-law field star distribution function.
\S\ref{sec4} summarizes our conclusions, discusses some of the limitations of our models, and how they might be improved in future work.

\section{Diffusion into the loss cone}
\label{sec2}
Traditional treatments of two-body relaxation focus on encounters that occur on timescales shorter than an orbit and neglect possible correlations between successive encounters \citep[e.g.][]{Chandrasekhar1942,Rosenbluth+1957}.
We call this non-resonant relaxation (NRR) to distinguish it from resonant relaxation (RR), as formulated by \cite{Rauch+Tremaine1996}, where the encounters between any two given bodies are correlated over many near-keplerian orbits, and their effects are averaged over the mean anomalies of both.

RR is a version of secular perturbation theory from celestial mechanics, but is applied to a system of very many minor bodies---in our case, stars.
In the secular approximation, minor bodies exchange angular momenta but not keplerian binding energies ($E=GM_\bullet/2a$), so that their semi-major axes ($a$) are unaffected.

A further distinction is made between vector and scalar RR:  the former averages also over the argument of pericenter, so that only the direction of each star's angular momentum (i.e., the orientation of its orbital plane) is affected; scalar RR does not make this second average. When we refer to RR, we intend the scalar variety, because we are interested in changes in eccentricity ($e$), or equivalently, normalized angular momentum
\begin{equation}\label{eq:jdef}
   j\equiv\sqrt{1-e^2}=J/\sqrt{GM_{\bullet}a}\,. 
\end{equation}

\begin{figure}
\figurenum{2}
\epsscale{1.15}
\plotone{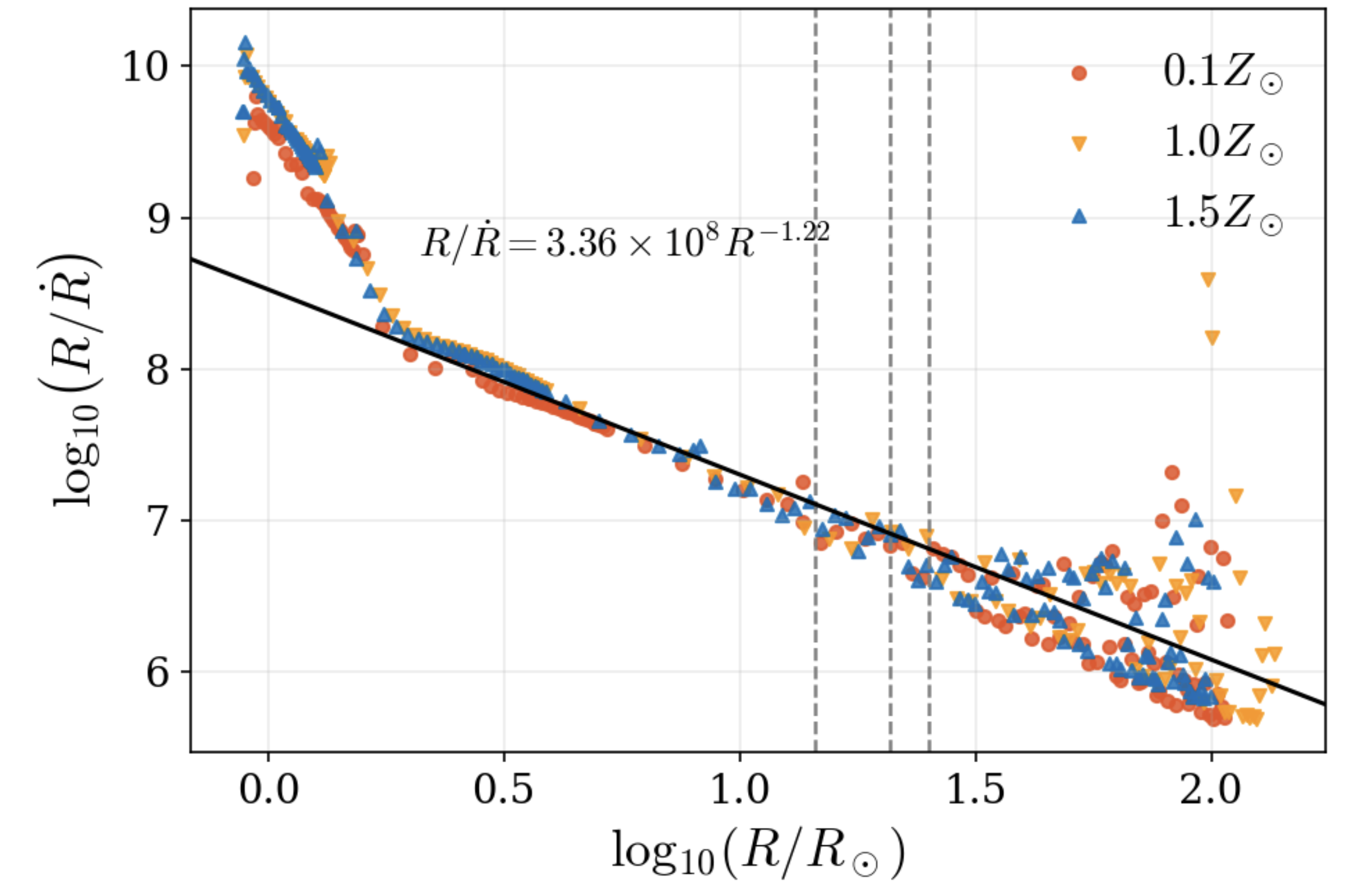}
\caption{Radius-dependent evolutionary lifespan scatter of various metallicity 1.0 $M_{\astrosun}$ stars as obtained through MESA, where $R/\dot{R}$ is given in years. The rest of the configurations are as given in Fig. (\ref{fig1}). The line of best fit is calculated from all points $\log_{10}(R)\geq 0.25$, with $R[R_{\astrosun{}}]$. Note that all radius-converted luminosity limits ($65 L_{\astrosun}$, 115$L_{\astrosun}$, 155$L_{\astrosun}$, in that order) are within the well-behaved region of the scatter.}
\label{fig2}
\end{figure}
 
\subsection{General Assumptions}

In all calculations and theoretical assumptions, we take ($r_{h}$) radius of SMBH influence ($r_{h}$) to be $2 \ \textrm{pc},$ where 
\begin{equation}
r_{h} = \frac{GM_{\bullet}}{\sigma^{2}(r_{h})}.
\end{equation}

We take the mass of Sgr A* to be $M_{\bullet} =4.3 \times 10^{6}\,M_{\astrosun}$ \citep{gravity}. We also assume that the field stars are Sun-like and non-evolving, with number density $n(r) \propto r^{-\gamma}$ in three dimensions and isotropic velocities, whence their distribution function $f \propto E^{\gamma-1.5}$.

We consider $\gamma=1.25$ and $\gamma=1.75$.  The latter corresponds to an idealized Bahcall-Wolf cusp, while the former approximates the shallower distribution of the fainter ($K>16$) visible stars \citep{schodel20}. 
\\
\subsection{Non-Resonant Relaxation Coefficients}

Following \cite{balex16} and Appendix~L of \cite{bitre08},
we convert the local velocity diffusion coefficients into the local diffusion coefficients for energy and angular momentum, then orbit average. The explicit NRR coefficients are provided in Appendix~B.

\subsection{Resonant Relaxation Coefficients}
The methodology behind this section is taken from \cite{bar18}. We use \texttt{scRRpy} \citep{bar19} to calculate the second-order diffusion coefficient $D_{jj}^{RR}$ using spherical harmonics up to degree $l=10$.
Convergence tests indicate that the results are sufficiently accurate for our purposes.

Expressing the Hamiltonian of the test star in terms of a mean potential and a fluctuating one, we have that
\begin{equation}\label{eq:hamiltonian}
H= \Phi_{K}(a)+\Phi_{GR}(a,J)+\Phi_{\star}(a,J) + H_{1}(a,J,J_{z},t),
\end{equation}
where $\Phi_{K}(a)$ is the standard keplerian gravitational potential, $\Phi_{GR}(a,J)$ is the leading-order relativistic correction to the potential, $\Phi_{\star}(a,J)$ is the mean potential due to the stars, and $H_{1}(a,J,J_{z},t)$ is the fluctuating-potential term. Neither $\Phi_{GR}$ nor $\Phi_{\star}$ directly affects the eccentricity or angular momentum, but both contribute to apsidal precession, thus limiting the correlation time of resonant relaxation \citep{Rauch+Tremaine1996}.

Not included here (or in \texttt{scRRpy}) is a term $\Phi_\mathrm{tide}$ to represent the contribution to the two-body potential due to the tidal distortion of the star.  As with close binary stars, this distortion also contributes to apsidal precession in the same sense (i.e., prograde) as $\Phi_{GR}$.  Using standard secular perturbation theory, we find that the ratio of the two contributions to the apsidal precession rate for highly eccentric orbits is
\begin{align}\label{eq:precession}
    \frac{\dot\varpi_\mathrm{tide}}{\dot\varpi_\mathrm{GR}}&\approx\frac{105}{256}k_2 \left(\frac{M_*}{M_\bullet}\right)^{1/3}\frac{c^2R_*}{GM_\bullet}\left(\frac{r_t}{r_p}\right)^4\nonumber\\
    &\lesssim \left(\frac{k_2}{0.285}\right)\left(\frac{R_*}{150\,R_{\odot}}\right)\left(\frac{M_*}{M_\odot}\right)^{1/3}\left(\frac{r_t}{r_p}\right)^4\,,
\end{align}
where $r_t\equiv R_*(M_\bullet/M_*)^{1/3}$ is the tidal disruption radius\footnote{We take the slightly more conservative value without the factor of $\sqrt[3]{2}$, which marginally justifies the RG envelope stripping assumption below.}, $r_p\equiv a(1-e)\ll a$ is the keplerian pericenter, and $k_2$ is the quadrupolar Love number of the red giant, which we've scaled by its value for a coreless $n=1.5$ polytrope. In fact, for $1\,M_\odot$, $Z=0.02$, the maximum of $k_2$ on the RGB is $\approx0.12$, declining to $\approx0.040$ at the RGB tip (MESA).

All of the terms in Eq. (\ref{eq:hamiltonian}) are evaluated in the secular approximation: i.e., they are averaged over the keplerian orbits (mean anomalies) of the test and field stars.  But $H_1$ remains time dependent because of slow changes in the Runge-Lenz vectors of the field stars.

Note that if the total mass of the background stars is kept constant, scaling the individual masses of the field stars  (and reciprocally scaling their number) has approximately the same effect on both the RR and NRR diffusion coefficients. 

\subsection{Fokker-Planck Time Evolution}

\subsubsection{Mean Exit Time}
A simple application of the diffusion coefficients is to see how long it takes, on average, for a star in orbit to reach the normalized angular momentum below which the star is tidally disrupted \citep{tidal},
\begin{equation}
    j_{t}=\sqrt{1-\Big(1-\frac{r_{t}}{a}\Big)^{2}}\approx\sqrt{\frac{2r_t}{a}},
\end{equation}
or that below which it plunges directly into the black hole \citep{Frank+Rees1976}, neglecting spin:
\begin{equation}
j_{\mathrm{plunge}}(a)=\frac{4}{c}\sqrt{\frac{GM_{\bullet}}{a}}.
\end{equation}

The ratio $j_t/j_\mathrm{plunge}$ depends upon the mean density of the star, the mass of the black hole, and the dimensionless factor $f$ in the scaling $r_t=fR_*(M_\bullet/M_*)^{1/3}$.  In this paper, we take $f=1$.  In that case, $j_t>j_\mathrm{plunge}$ for all late-type stars of interest here, certainly on the RGB but even---at least marginally---on the main sequence.

With our assumptions, the loss cone $j\le j_\mathrm{min}$ should be empty at $a\ll 1\unit{pc}$.
That is to say, orbital periods are short compared to the diffusion time across the loss cone, $t_\mathrm{diff}\sim j_\mathrm{min}^2/D_{jj}$.
Therefore, once a star reaches $j_{\mathrm{min}}=j_t$, we assume that the star is effectively lost.

We have, however, set the loss cone where tidal stripping becomes important, and one may question whether red giant envelopes can be stripped so completely in a single passage as to prevent the star from continuing on the RGB.
This assumption deserves scrutiny, but for our present purposes, it provides an upper bound to the effectiveness of tidal encounters in accounting for the dearth of bright red giants.

In order to derive the mean exit time of the stars starting from an arbitrary position $j_{0}$ in $j$-space at time $t=0$, we use the (forward) Fokker-Planck equation \citep{balex14} to describe the evolution of the probability density $P(j,t)$ that the test star has normalized angular momentum $j$ at time $t$: 
\begin{multline}
\frac{\partial P(j,t)}{\partial t} = -\frac{\partial}{\partial j}D_{j} P(j,t) + \frac{1}{2}\frac{\partial^{2}}{\partial j^{2}}D_{jj} P(j,t)\\
\equiv\mathcal{L}P(j,t)\,,
\label{eq:Pevol}
\end{multline}
where we've introduced the shorthand $\mathcal{L}$ for the differential operator on the righthand side. Our earlier assumption that diffusion starts at $j=j_{0}$ is equivalent to 
\begin{equation}
P(j,0)=\delta (j-j_{0}),
\end{equation}
and the boundary conditions on $P(j,t)$ are 
\begin{equation}
P(j',t'|j_{\mathrm{min}},t)=\partial_{j}P(j',t'|j,t)\Big\vert_{j=1}=0,
\end{equation}
where the absorbing boundary at $j=j_{\min}$ yields the first vanishing term and the reflective boundary at $j=1$ yields the second (detailed in \cite{gard04}), although $D_{jj}(1)=0$ regardless. 

Assuming that the field star distribution is isotropic, 
a thermal distribution $P(j)=2j$ should be a steady-state solution to eq.~\eqref{eq:Pevol} in the absence of the loss cone. Therefore \citep{balex16}
\begin{equation}
2jD_{j}=\partial_{j}(jD_{jj}).
\end{equation}

The probability that the star has not yet been lost (fallen into the loss cone) up to time $t>0$, given that it started from $j$ at $t=0$, is
\begin{equation}
G(j,t)= \int_{j_\mathrm{min}}^{1} P(j',t | j'(0)=j,0) \,dj'.
\label{eq:Gdef}
\end{equation}

The equivalent boundary conditions on $G(j,t)$ are that it should vanish at the loss cone, $G(j_\mathrm{min},t)=0$, and that its derivative $\partial_j G(j,t)=0$ at $j=1$. 
We take $D_{j}$, $D_{jj}$, and the boundary conditions at $j=j_\mathrm{min}$ and $j=1$ to be constant in time. The former presumes that the field-star distribution is statistically stationary, while the latter ignores changes in the mean density of the test star as it evolves up the RGB. 
The system is then time-homogeneous, whence
\begin{equation}\label{eq:homogeneous}
P(j',t |j,0)= P(j',0 |j,-t).
\end{equation}

It can then be shown that $G(j,t)$ satisfies the backward Fokker-Planck equation [BFPE; see \citealt[\S5.2.144]{gard04}]
\begin{multline}
\frac{\partial G(j,t)}{\partial t} = D_{j}(j,t)\frac{\partial}{\partial j}G(j,t) + \frac{1}{2}D_{jj}(j,t)\frac{\partial^{2}}{\partial j^{2}}G(j,t)\\
\equiv\mathcal{L}^\dagger G(j,t).
\label{eq:Gevol}
\end{multline}

As the notation suggests, the differential operators in eqs.~\eqref{eq:Pevol} \& \eqref{eq:Gevol} are adjoints.
We can then define the mean exit time (MET) as follows:
\begin{equation}
T(j)=-\int_{0}^{\infty} \frac{\partial G(j,t)}{\partial t}\  t \ dt.
\label{eq:Tdef}
\end{equation}

In fact, the probability density~\eqref{eq:homogeneous} itself satisfies the BFPE, with the understanding that $\mathcal{L}^\dagger$ acts on $j$ (not $j'$) in that function.
Using this fact and the definitions \eqref{eq:Gdef} \& \eqref{eq:Tdef}, it can further be shown that
\begin{equation}
\mathcal{L}^{\dagger}T(j) = -1.
\label{Eq5}
\end{equation}

Equation~\ref{Eq5} is an ordinary differential equation and can be integrated:
\begin{equation}
T(j)= 2\int_{j_{\mathrm{min}}}^{j} \frac{dj'}{\psi(j')}\int_{j'}^{1} \frac{\psi(j'')}{D_{jj}(j'')} \ dj'',
\label{Eq10}
\end{equation}
where
\begin{equation}
\psi(j)= \exp \int_{j_{\mathrm{min}}}^{j}dj'\frac{2D_{j}(j')}{D_{jj}(j')}.
\end{equation}

\begin{figure}
\figurenum{3}
\epsscale{1.20}
\plotone{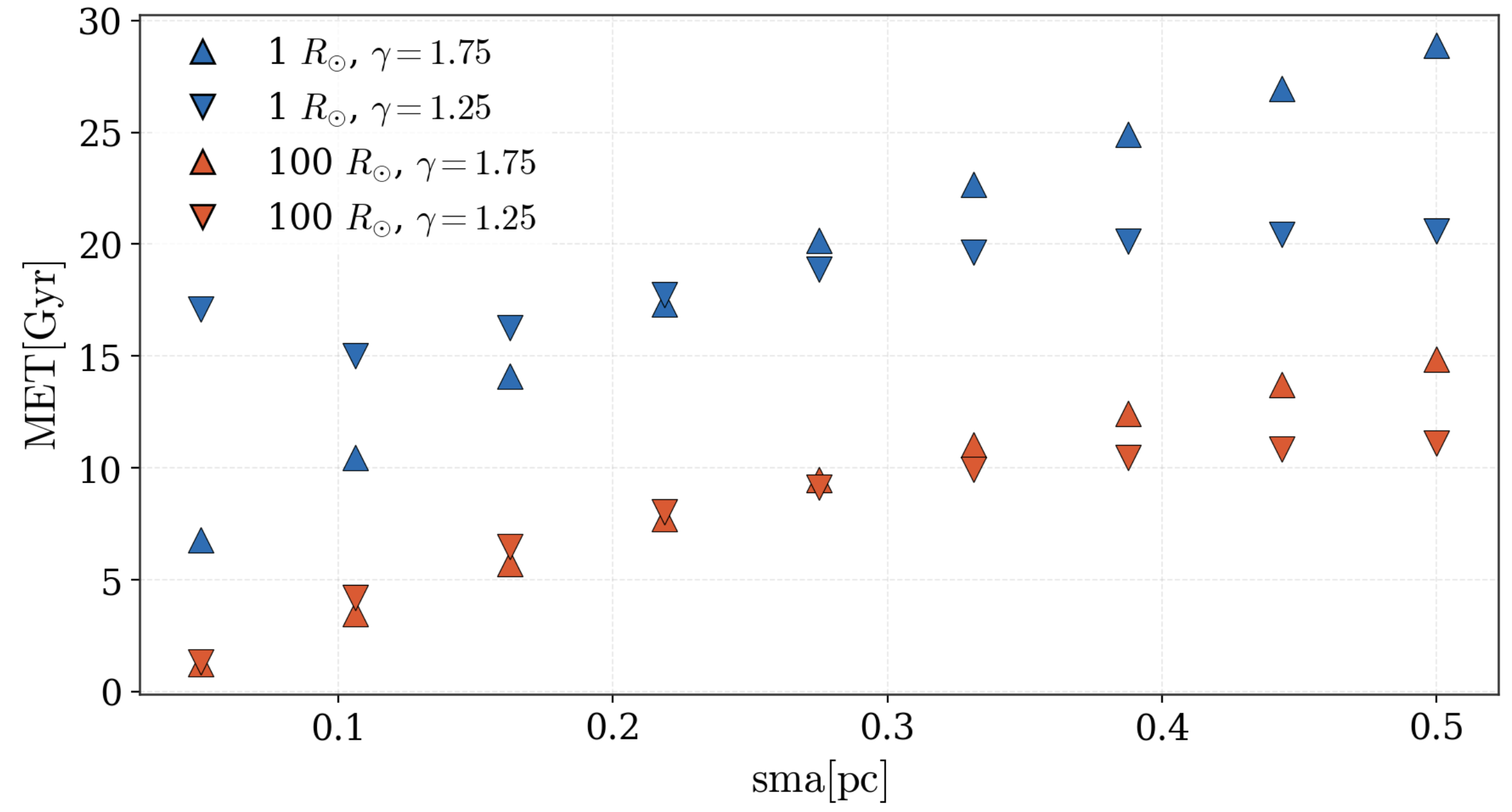}
\caption{Illustration of the dependence of mean exit time on stellar radius and on the power-law index of the density profile of field stars, given a test star with solar mass, where $j_{0}=0.7$.}
\label{fig3}
\end{figure}

Applying the derived relation to the assumed model yields Fig. (\ref{fig3}), the corresponding METs. We vary the radius of the escaping stellar body, indicated by the varying colors. We assume that the star starts diffusing from $j=0.7,$ the median value for isotropically distributed orbits in a keplerian potential. 

Figure (\ref{fig3}) depicts METs mostly decreasing as the stellar radius increases.
However, even at $a\sim 0.1\,\mathrm{pc}$, the MET is much longer than the time remaining on the RGB at $100\,\mathrm{R_\odot}$ ($\ll10^9\,\mathrm{yr})$.
Furthermore, the MET is a weak function of stellar radius, and therefore does not discriminate strongly between bright red giants and main-sequence stars.

This should not be surprising: while the size of the tidal disruption loss cone \citep{tidal} is roughly proportional to stellar radius, the rate at which stars diffuse into the loss cone depends on its size only logarithmically \citep{Frank+Rees1976}---at least insofar as NRR dominates.
If resonant relaxation were entirely dominant, one could imagine a situation in which relativistic apsidal precession protected main-sequence stars, but not red giants, from reaching sufficiently eccentric orbits to be disrupted. However, for the parameters considered here, NRR and RR are of comparable importance.

\subsection{Monte Carlo Diffusion Simulations}
\label{mcds}
Monte-Carlo simulations have been conducted to ensure the validity of the methodology above \citep[e.g.][]{balex16}. It also facilitates the incorporation of additional physical effects: in particular, the evolution of stellar radius on the RGB and stellar collisions.

From \cite{risken}, we simulate the diffusion of a test star in random increments $\Delta j$ at each $\Delta t$ timestep, taken as 0.1 kyr \citep{tep21}:
\begin{equation}
\Delta j = D_{j}(j)\Delta t+\sqrt{D_{jj}(j) \Delta t}\  \xi(t). 
\label{Eq16}
\end{equation} 

Here, $\xi(t)\sim\mathcal{N}(0,1)$ represents samples from the unit normal distribution, chosen independently at each time step.
The contributions from NRR and RR are chosen independently and summed \citep{tep21}.
We find that when the test-star radius is held fixed (rather than evolving on the RGB) the MC method predicts METs that are consistent with those in Fig. (\ref{fig3}) up to statistical fluctuations.

\begin{figure}
\figurenum{4}
\epsscale{1.15}
\plotone{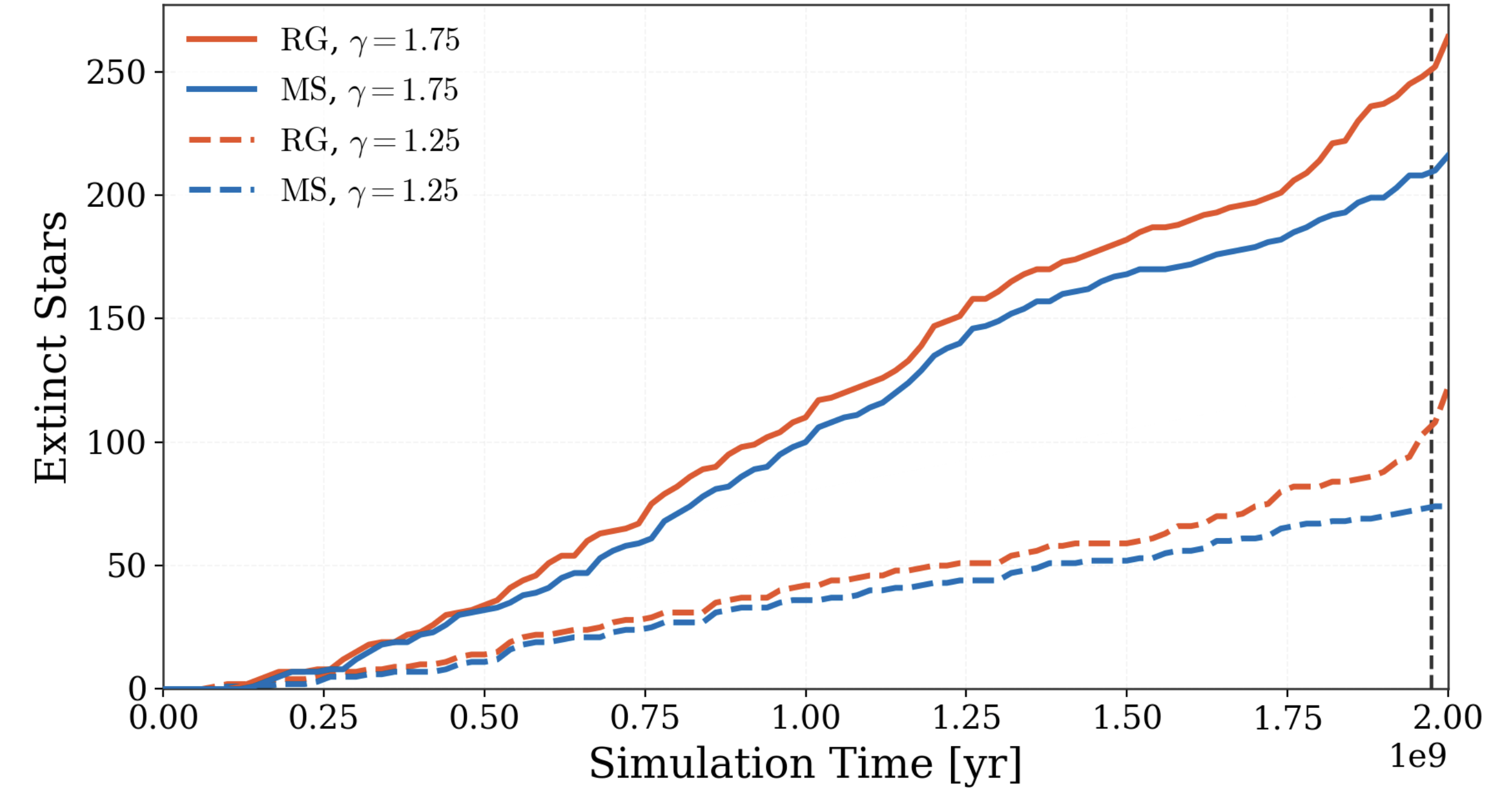}
\caption{Stars lost by diffusion into the loss cone from an initial population of $10^{3}$ Sun-like ($M_*=1\,M_\odot$, $Z=0.02$) test stars, simulated by Monte-Carlo methods  over 2 Gyr on the main sequence (\emph{blue}) and the sub-giant and red-giant branches (\emph{red}). In all cases, $j_0=0.7$, $a=0.1$ pc, $\Delta t = 0.1$ kyr. The dotted vertical line denotes the time corresponding to $115L_{\astrosun}$.}
\label{fig4}
\end{figure}

The results of the simulation for main-sequence stars and evolving red giants are shown in Fig.~\ref{fig4} and Table~\ref{tab1}. Although we do see more red giants than main-sequence stars lost by diffusion into the loss cone (where they are tidally disrupted), the differences are probably too small to explain the observations, especially when one considers that they would have been even smaller had we compared bright red giants to fainter ones, rather than to main-sequence stars, since the size of the loss cone is a function of mean density.

\begin{table}[htbp]
\footnotesize                 
\setlength{\tabcolsep}{5pt}   
\begingroup
\renewcommand{\arraystretch}{1.25} 
\noindent
\begin{flushleft}
\begin{tabular}{lcccccc}
\hline
\multicolumn{2}{l}{} & \multicolumn{4}{c}{semimajor axis[pc]} \\
\cline{3-6}
Group & $\gamma$ & 0.05 & 0.10 & 0.25 & 0.50 \\
\hline
RG & 1.75 & $28.4 \pm 1.7$ & $26.7 \pm 0.9$ & $15.7 \pm 1.2$ & $13.3 \pm 1.9$ \\
   & 1.25 & $6.1 \pm 0.6$ & $13.0 \pm 0.9$ & $12.2 \pm 1.2$ & $11.7 \pm 1.0$ \\
\hline
MS & 1.75 & $18.8 \pm 1.9$ & $21.0 \pm 2.5$ & $13.7 \pm 0.8$ & $13.1 \pm 2.0$ \\
   & 1.25 & $3.6 \pm 0.2$   & $8.6 \pm 0.6$  & $9.7 \pm 1.1$  & $10.3 \pm 0.7$ \\
\hline
\end{tabular}
\end{flushleft}
\endgroup
\vspace{1pt}
\caption{Extinction percentages and standard deviation uncertainties induced by resonant and non-resonant diffusion alone, without stellar collisions.}
\label{tab1}
\end{table}
\section{Stellar Two-Body Collisions}\label{sec3}
In this section, we add stellar collisions to the Monte-Carlo simulations. 
The envelopes of bright red giants are particularly susceptible to collisions because of their large geometric cross section and low surface escape velocity.
Note from Fig.~(\ref{fig2}) that the evolutionary timescale on the RGB declines more rapidly with stellar radius than $R_*^{-1}$ but more slowly than $R_*^{-2}$.
The collision rate scales as $R_*$ in the gravitational-focusing regime, where $\sigma\lesssim \sqrt{GM_*/R_*}$, and as $R_*^2$ in the geometric regime.
With a power-law number density ($\propto r^{-\gamma}$) in a potential dominated by the black hole, the 1D velocity dispersion $\sigma= \sqrt{GM_\bullet/(\gamma+1)r}$, which comes to $260\unit{km\,s^{-1}}$ at $r=0.1\unit{pc}$, whereas $\sqrt{GM_\odot/R_\odot}\approx440\unit{km\,s^{-1}}$.
Thus, while gravitational focusing is somewhat important for the main-sequence stars, the geometric cross section dominates on the RGB.
Consequently, since the cross section increases with radius more rapidly than the lifetime decreases, most of the collisions experienced by a red giant should occur at the most advanced stages of its evolution.
This is very promising for explaining the paucity of bright giants compared to faint ones.

In examining these collisional effects, we make some broad assumptions. 

\subsection{Collision Rates}
We assume that collisions at pericenters $\lesssim R_*$ are completely destructive.
As noted in \S\ref{sec:intro}, \cite{rose23} allowed for the possibility of mergers among main-sequence stars, but this is less likely at $r\lesssim0.1\unit{pc}$ if one of the collision partners is a red giant because of the smaller binding energy of its envelope \citep{Bailey+Davies1999}.

We assume a keplerian relative potential\footnote{Following \cite{bitre08}, the relative potential $\Psi(r)$ and relative energy $E \equiv \Psi - v^2/2$ are minus the usual potential and energy, hence positive for bound stars.}
$\Psi(r) = GM_\bullet/r$, 
and a power-law isotropic distribution function for the field stars,
\begin{equation}
\label{F}
f(E) = f_0 E^{\gamma - 3/2}\qquad (f_0,\gamma=\text{constants}),
\end{equation}
which implies\footnote{Equivalently, $n(r)=n_{0}r^{-\gamma}, n_{0}=10^{6}(\frac{M_{\astrosun}}{m_{f}})\cdot \frac{3-\gamma}{4\pi}(1\textrm{pc})^{\gamma-3},$ following \cite{schodel18}.}
\begin{equation}\label{eq:n_of_r}
n(r) = f_0(2\pi)^{3/2}\frac{\Gamma(\gamma-\tfrac{1}{2})}{\Gamma(\gamma+1)}
\left(\frac{GM_\bullet}{r}\right)^\gamma.
\end{equation}

The collision cross section is the sum of geometric and gravitational-focusing terms:
\begin{equation}\label{eq:sigma}
\sigma(v_{\mathrm{rel}})=A+\frac{B}{v_{\mathrm{rel}}^2},
\end{equation}
with $A=\pi r_p^2$ and $B=2\pi G(m+m')r_p$ for stars of masses $m,m'$ on a hyperbolic relative orbit with pericenter $\lesssim r_p$, for which we take the radius $R_*$ of the test star. The local collision rate between a test particle with velocity $\boldsymbol{v}'$ and field stars with isotropic velocity distribution $f(v)$ is
\begin{equation}\label{eq:C_r}
C(r)=\int f(v)\,|\boldsymbol{v-v}'|\,\sigma(|\boldsymbol{v-v}'|)\,d^3\boldsymbol{v}.
\end{equation}

The calculation of $C(r)$ and its orbit average for a power-law distribution of field stars is detailed in Appendix A.
This is similar to \cite{hypergeometric}, but those authors replace $v_\mathrm{rel}$ with the local velocity dispersion $\sigma(r)$, whereas we use the more exact Eq.~\eqref{eq:C_r}.

\subsection{Collisional Simulations}
We conducted similar Monte-Carlo simulations for the same parameters used in \S\ref{mcds}. 
Results are shown in Table~\ref{tab:collisions_clean}.
Two sets of results are shown for the red giants: in columns 5 \& 6, the evolution was halted at a luminosity corresponding roughly to $K_s=15$, while the last two columns went all the way to the helium flash.

\begin{table*}
\small
\setlength{\tabcolsep}{6pt}
\renewcommand{\arraystretch}{1.0}
\noindent
\begin{flushleft}
\begin{tabular}{ccccccccccc}
\hline
$\gamma$ & sma [pc] & \multicolumn{2}{c}{Main Sequence} & \multicolumn{2}{c}{RG ($115 L_{\astrosun}$) }  & \multicolumn{2}{c}{RG (HF)} \\
\cmidrule(lr){3-4} \cmidrule(lr){5-6} \cmidrule(lr){7-8}
 &  & LC & Col. & LC & Col. &  LC & Col.  \\
\hline
1.75 & 0.05 & $170 \pm 13.3$ & $199 \pm 13.3$ & $146 \pm 12.5$ & $681 \pm 7.9$ &	$124	\pm 10.4$&	$868	\pm 10.7$\\
     & 0.10 & $198 \pm 13.0$ & $49 \pm 5.0$& $214 \pm 12.5$  & $291 \pm 6.1$ &$220 \pm	13.1$	&$571	 \pm 15.7$\\
     & 0.25 & $145 \pm 7.4$   & $18 \pm  4.0$ & $157 \pm 6.7$& $55.3 \pm  4.6$&	$145 \pm	11.1$&	$205	 \pm 12.8$\\
     & 0.30 & $114 \pm 7.0 $  & $15 \pm 3.5$ & $123.7 \pm  5.0$& $45 \pm 6.7$&$146 \pm	11.2$&	$137	 \pm 10.9 $\\
     & 0.36 &$ 108 \pm 7.0$  & $10 \pm 3.3$ & $118 \pm 7.4 $& $37.3 \pm 3.3$ & $117 \pm 10.2$ & $99\pm 9.4$ \\
\hline
1.25 & 0.05 & $27.3 \pm 2.9$   & $49 \pm 5.5$ & $33.7 \pm 5.3$& $266 \pm 5.6$ &$35	\pm5.8$&	$719	\pm14.2$ \\
     & 0.10 & $87.7 \pm 15.6$ & $19 \pm 2.5$ & $116.3 \pm 20.0$ & $97.3 \pm 7.4$ &$108	\pm9.8$&	$318\pm	14.7$ \\
     & 0.25 & $101  \pm 7.9$  & $8 \pm 1.0$  & $116.3 \pm 3.8$& $30.3 \pm 4.7$ &$109	\pm9.9$	&$96	\pm9.3$\\
     & 0.30 & $103  \pm 7.3$  & $6 \pm 1.0$ & $117 \pm 12.6$ & $26.7 \pm 6.6$ &$130	\pm 10.6$	&$65	\pm7.8$ \\
     & 0.36 & $80.6 \pm 2.6$    & $6 \pm 1.0$ & $93.3 \pm\ 4.0$& $15.7 \pm 4.5$ &	$118	\pm 10.2$&	$56	\pm7.3$\\
\hline
\end{tabular}
\end{flushleft}
\vspace{2.5pt}

\caption{MC simulations, each for $10^3$ stars of $1\,M_\odot$ under the joint influence of diffusion into the loss cone (LC) and stellar collisions (Col.). Columns 3-8 list estimated mean and standard deviation in the number of extinct stars (out of $10^3$). Cols. 3-4: main-sequence stars with fixed radii followed for 2~Gyr.  Cols. 5-6: red giants followed from the end of the main sequence up to $115 L_{\astrosun}$. Cols. 7-8: similarly, but up to the helium flash ($L\approx 2800\,L_\odot$, $R_*\approx 180\,R_\odot$).}
\label{tab:collisions_clean}
\end{table*}

\begin{figure}
\figurenum{5}
\epsscale{1.15}
\plotone{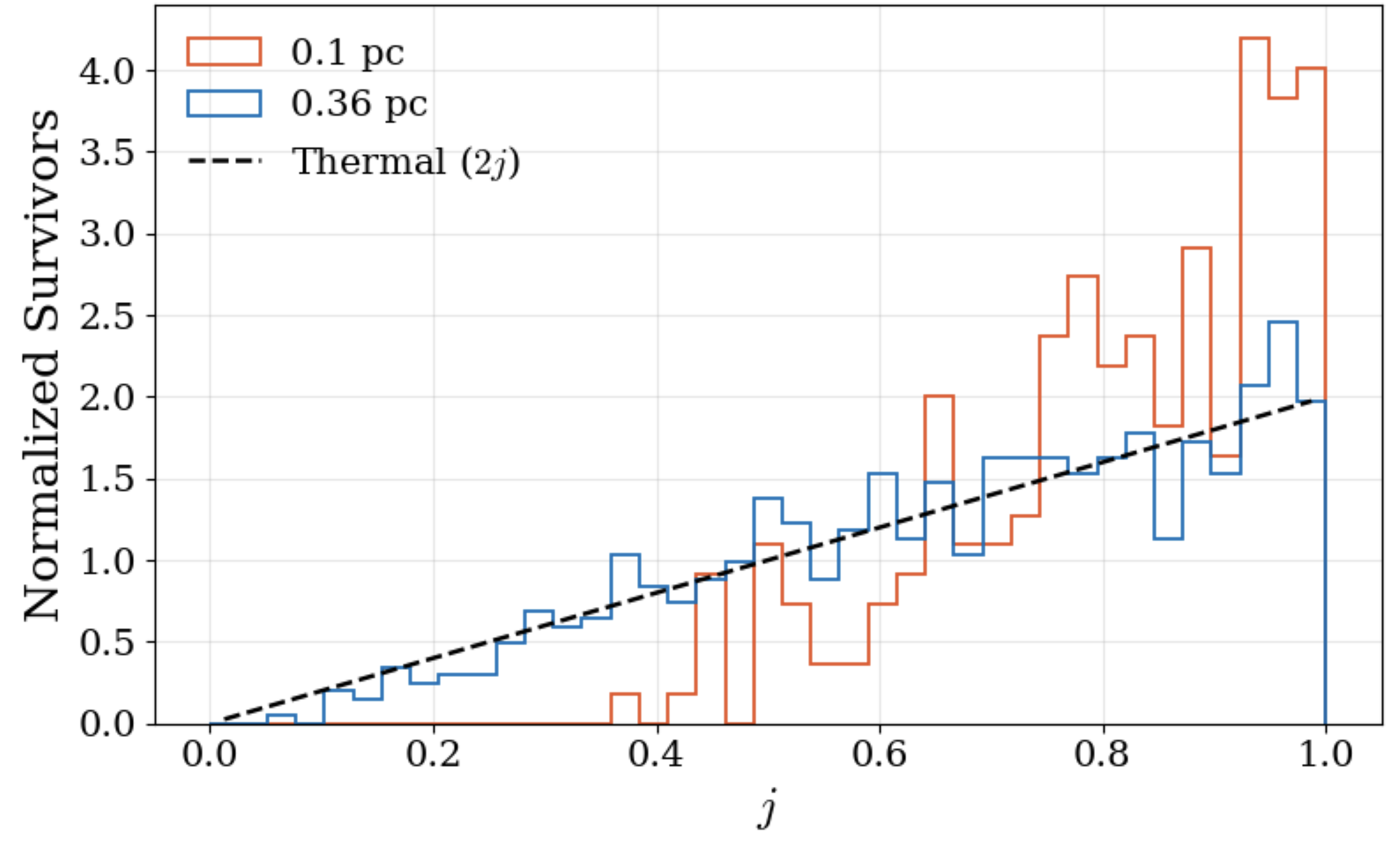}
\caption{The normalized distribution in dimensionless angular momentum $j\equiv\sqrt{1-e^2}$ of $1\,M_\odot$ red giants not destroyed until the helium flash. Simulations as in Table~\ref{tab:collisions_clean} with $\gamma=1.75$, but for semimajor axes $a=0.1\unit{pc}$ (\textit{red}) and $0.36\unit{pc}$ (\textit{blue}).}
\label{fig5}
\end{figure}

The outcomes are categorized in these tables by ``cause of death": diffusion into the tidal-disruption loss cone versus collisions.
As before (\S\ref{sec2}), diffusion scarcely distinguishes the red giants from the main-sequence stars.
As can be seen, the red giants are much more vulnerable to collisions, as expected.
Furthermore, comparison of columns 6 \& 8 of Table~\ref{tab:collisions_clean} shows that the red giants suffer significantly more collisions \emph{after} they reach $115\,L_\odot$, at least for semimajor axes $\ge 0.1\unit{pc}$ (below which the majority are destroyed even before they reach $115\,L_\odot$).  This shows the importance of the geometrical term ($\propto R_*^2$) in their collision cross sections. 

Figure~\ref{fig5} plots the angular momenta of the red giants that are not destroyed, for two values of the semimajor axis (which is conserved throughout the evolution, as we do not allow for energy diffusion). First, since all of the stars started at $j_{0}=0.7$, it can be seen that diffusion in angular momentum is quite effective over 2~Gyr. The final distribution in $j$ is almost thermal (i.e., uniform in phase space) at the larger semimajor axis ($a=0.36\unit{pc}$), where most of the stars survive (Table~\ref{tab:collisions_clean}). Secondly, at the smaller value of $a$ (where few stars survive the entire RGB), the survivors tend to have less eccentric orbits.

This is due not mainly to the loss cone, but to collisions, and in particular to the geometric part of the collision cross section. The gravitational-focusing part is much less sensitive to eccentricity. In fact, the orbit-averaged contribution from gravitational focusing is completely independent of eccentricity for $\gamma=3/2$, whereas the geometric contribution is proportional to $j^{-1}=1/\sqrt{1-e^2}$ (Appendix~A).
Orbital diffusion is unable to fully isotropize the survivors at $a=0.1\unit{pc}$ because most of the collisions occur on the upper part of the RGB, where the evolutionary lifetime is $\lesssim10^8\unit{yr}.$

An implication for observations is that the velocity dispersion of the red giants at $r\lesssim5''$ in projection should be anisotropic if their destruction is due to collisions.  This could be tested by analyzing proper motions, but has not yet been, as far as we know.

\section{Conclusion and Future Plans}\label{sec4}
As reviewed in \S\ref{sec:intro}, there is a dearth of red giants relative to fainter---but probably still post-main-sequence---stars in the inner $0.1-0.5\unit{pc}$ of the Galactic Center.
We conclude that this cannot be explained solely by orbital diffusion into the loss cone, where stars are tidally stripped or swallowed whole by the central black hole, because the rate of diffusion into the loss cone depends only logarithmically on stellar radius ($R_*$), whereas the lifetime remaining on the red-giant branch (RGB) declines as a power (greater than unity) of $R_*$.
The inclusion of resonant scalar relaxation of eccentricities, as well as conventional two-body relaxation, does not change this conclusion, at least insofar as both effects are well described as diffusive processes.

Interestingly, the Monte-Carlo simulations predict a tangential anisotropy in the eccentricity distributions of the late-type bright stars closer to Sgr A* ($a\lesssim 0.1$ pc), which eventually thermalizes further from Sgr A*. We plan on examining this further, with the goal of deciding whether the anisotropy should be detectable in future observations of the late-type stars, or even in published observations \citep[e.g.][]{Schoedel+2009}.

For simplicity, we have taken all of the stars---test and field---to have the same mass, namely $1~M_\odot$.
As we have argued, this is likely a fair approximation for the late-type stars bright enough to be counted in the $K_s$ band, because only stars of roughly that initial mass would be on the RGB (or red clump or AGB) today if they formed $\sim10\unit{Gyr}$ ago.
But it is surely not true of the field stars that perturb the red-giant's orbits or collide with them.
The old field stars and stellar remnants might be expected to have relaxed into a steady-state Bahcall-Wolf cusp \citep[but see][]{Merritt2010}, in which case mass segregation would cause the heaviest objects to have a number-density profile at least as steep as $r^{-1.75}$ \citep{Bahcall+Wolf1976,Bahcall+Wolf1977,fpmassseg}.
This is steeper than the deprojected slope of the observed counts in the range $18\lesssim K_s<15$ \citep{Gallego-Cano+2018,schodel20}.

Thus, if the old field star distribution has indeed reached a steady state, it is probably dominated at $r\lesssim0.1\unit{pc}$ by unseen degenerate remnants that are individually heavier than the visible stars: e.g., neutron stars and stellar-mass black holes.
In that case, holding the mean mass density of the field stars constant but taking $m_f>1$ as a characteristic value for their individual masses relative to that of the Sun, we have probably underestimated both the resonant and non-resonant diffusion coefficients by a factor $\sim m_f^{-1}$, to the extent that the local stellar mass density is fixed.

On the other hand, we may have slightly overestimated the effective collision rates.
If a grazing collision with a solar mass field star at pericenter $r_p\le R_*$ is sufficient to do substantial damage to a solar-mass red giant of radius $R_*$, then the relevant pericenters for field stars of mass $m_f\times(1\,M_\odot)$, with $m_f>1$, would be $r_p\le m_f^{1/3} R_*$.
The rate of such collisions scales with $m_f$ as $m_f^{-1/3}$ in the geometric regime (the relevant regime for the giants) since the number density of the field stars should scale as $m_f^{-1}$.
So ours is a rather modest overestimate of the effective collision rate even if stellar-mass black holes dominate ($m_f\sim10$).
    
We assumed that any collision of the test star with a field object resulted in the complete destruction of the test star. This is probably our most questionable simplification.  If the mass remaining in the envelope after a single such collision is less than the difference between the mass of current helium core ($\approx0.28\,M_\odot$ at $L=115\,L_\odot$)
and the mass of the core at the helium flash ($\approx0.47\,M_\odot$), the envelope will quickly re-expand to almost the same photospheric radius and luminosity, and will have enough mass complete its evolution up the RGB.
Therefore, either a deeply penetrating collision, or a series of less violent ones, will probably be needed to halt the evolution at an observed magnitude $K_s\ge15$.
Quantifying the meaning of ``deeply penetrating" will require careful hydrodynamic simulations beyond the scope of the present work.

\goodbreak
\appendix{}
\section{Orbit-averaged collision rate}
Writing the angular part of the integral \eqref{eq:C_r} explicitly, with $\theta$ the angle between $\boldsymbol{v}$ and $\boldsymbol{v}'$,
\begin{equation}\label{eq:C_r_expanded}
C(r)=2\pi\int_0^{\sqrt{2\Psi}} f(v)\,v^2\,dv
\int_{-1}^{1} d(\cos\theta)
\Bigg(A\sqrt{v^2+v'^2-2vv'\cos\theta}
+\frac{B}{\sqrt{v^2+v'^2-2vv'\cos\theta}}
\Bigg).
\end{equation}

Let $v_\downarrow=\min(v,v')$ and $v_\uparrow=\max(v,v')$. The integrations over $\cos\theta$ are elementary, yielding
\begin{equation}\label{eq:inner_integral}
\frac{2A}{3v_\uparrow}(3v_\uparrow^2+v_\downarrow^2)+\frac{2B}{v_\uparrow}.
\end{equation}

Because the identities of $v_\uparrow$ and $v_\downarrow$ depend on whether $v<v'$ or $v>v'$, the outer integral in \eqref{eq:C_r_expanded} must be split into the ranges $0\le v<v'$ and $v'\le v\le\sqrt{2\Psi}$. We handle the simpler $B$-term first. The contribution to $C(r)$ proportional to $B$ is
\begin{equation}\label{eq:CB_v}
C_B(r)=\frac{4\pi B}{v'}\int_0^{v'} f(v)v^2\,dv+4\pi B\int_{v'}^{\sqrt{2\Psi}} f(v)v\,dv\\
=\frac{4\pi B}{v'}\int_{E'}^{\Psi} f(E)\sqrt{2(\Psi-E)}\,dE
+4\pi B\int_0^{E'} f(E)\,dE,
\end{equation}
where we have changed variables using $E=\Psi-v^2/2$ and $E'\equiv\Psi-v'^2/2$. Introducing the antiderivative
\begin{equation}
\label{eq:F_of_E}
F(E)=\int f(E)\,dE = \frac{f_0}{\gamma-\tfrac{1}{2}}E^{\gamma-\tfrac{1}{2}},
\end{equation}
and integrating the first integral in \eqref{eq:CB_v} by parts gives the compact form
\begin{equation}
\label{eq:CB_final}
C_B(r)=\frac{4\pi B}{v'}\int_{E'}^{\Psi}\frac{F(E)}{\sqrt{2(\Psi-E)}}\,dE.
\end{equation}

By similar steps, one finds that the contribution $\propto A$ (the part involving the geometric cross section) becomes
\begin{equation}\label{eq:CA_explicit}
C_A(r)=4\pi A\Bigg[
v'\int_{E'}^{\Psi}\frac{F(E)}{\sqrt{2(\Psi-E)}}\,dE\\
+2\int_0^{E'} F(E)\,dE
+\frac{1}{v'}\int_{E'}^{\Psi}F(E)\sqrt{2(\Psi-E)}\,dE
\Bigg].
\end{equation}

The middle integral in \eqref{eq:CA_explicit} is elementary for the power-law $F(E)$ in \eqref{eq:F_of_E}. The other two integrals (one of which is the same as that appearing in \eqref{eq:CB_final}) are generally not elementary except for integral and half-integral $\gamma$, but they can be written in terms of incomplete beta functions. 

The local rates $C_{A}$ \& $C_{B}$ depend on radius---or equivalently,
$\Psi=GM_\bullet/r$---and $E'$, the test-particle energy.
The latter is of course constant along the test particle's orbit.
Furthermore, because the field-particle distribution function \eqref{F} and its
integral \eqref{eq:F_of_E} are taken to be power laws in $E$, the local collision rates $C_A$ and
$C_B$ can be written as powers of the test-particle's semimajor axis ($a$) times
dimensionless functions of $a/r$:
\begin{subequations}\label{scalings}
\begin{equation}
  \label{eq:scalingsa}
  C_A(r) \to A\times f_0(k/a)^{\gamma+\tfrac{1}{2}}g_A(a/r),
\end{equation}
\begin{equation}
  \label{eq:scalingsb}
C_B(r) \to B\times f_0(k/a)^{\gamma-\tfrac{1}{2}}g_B(a/r),
\end{equation}
\end{subequations}
with the abbreviation $k\equiv GM_\bullet$.
Note that $f_0(k/a)^{\gamma\pm\tfrac{1}{2}}$ have units of $n\times v^{\pm1}$.
So it suffices to calculate the orbital averages of these for a single value of the semimajor axis, since the averages for any other value of $a$ will follow from the scalings \eqref{scalings}.

We perform the orbit averages numerically by standard methods, expressing time and radius in the orbit in terms of eccentric anomaly. We note for debugging purposes that exact results are possible for $\gamma=3/2$: $\langle g_B\rangle=10\pi/3$, $\langle g_A\rangle = (\pi/15)\times (96/\sqrt{1-e^2}\,-17)$. 
One does have to make a separate computation for each orbital eccentricity of the test star, however.

Taking the number of collisions per timestep (in orbital periods), $N(a,e)$, we construct a Poisson distribution with mean $\mathbb{E}[N(a,e)]$. Drawing from the distribution will yield the probability of a collision-free timestep, as we consider any non-zero number of collisions equivalent to the destruction of the test star only. 

\section{Non-resonant-relaxation Diffusion Coefficients}
The NRR diffusion coefficients can be expressed in terms of the $F_{i}$ functions \citep{corud78}. We also employ Risken's \citeyearpar{risken} coordinate change to interpret the diffusion coefficients in terms of the normalized specific orbital angular momentum $j\equiv\sqrt{1-e^{2}}.$ The mass of the test star is denoted $M_{*}$ and the mass of individual field objects is denoted $m_{f}$.

  \begin{equation}
    D_{E}
      = -F_{0}
      + \frac{M_{*}}{m_{f}}\,F_{1}, 
      \quad
    D_{EE}
      = \frac{4E}{3}(F_{0}+F_{4}), 
      \quad
    D_{Ej}
      = \frac{2j}{3}(F_{4}-F_{5}),
    \label{Ecof}
  \end{equation}

  \begin{equation}
    D_{j}E
      = \frac{5-15j^{2}}{12j}\,F_{0}
      + \frac{M_{*}j}{2m_{f}}\,F_{1}
      - \frac{j(M_{*}+m_{f})}{2m_{f}}\,F_{2}
      + \frac{F_{3}}{j}
      - \frac{j}{6}\,F_{4}
      - \frac{j}{3}\,F_{5}
      - \frac{1}{3j}\,F_{7},
    \label{jcoff}
  \end{equation}

  \begin{equation}
    D_{jj}E
      = \frac{5(1-j^{2})}{6}\,F_{0}
      - \frac{j^{2}}{2}\,F_{2}
      + 2\,F_{3}
      + \frac{j^{2}}{3}\,F_{4}
      - \frac{2j^{2}}{3}\,F_{5}
      + \frac{j^{2}}{2}\,F_{6}
      - \frac{2}{3}\,F_{7}.
    \label{jcof}
  \end{equation}
\bibliography{bibliography}{}
\bibliographystyle{aasjournalv7}

\end{document}